\documentclass[letterpaper,12pt]{article}
\pdfoutput=1 
\usepackage[bottom]{footmisc} 
\usepackage{amssymb}
\usepackage{amsthm}
\usepackage{newtxmath}
\usepackage{microtype} 
\usepackage[utf8]{inputenc}
\usepackage{newtxtext} 
\usepackage{setspace}
\usepackage{lineno,xcolor}
\usepackage{graphicx}
\usepackage{float}
\usepackage{printlen}
\usepackage{relsize}
\usepackage[hidelinks]{hyperref}
\usepackage{pdfpages}
\usepackage{makecell} 
\usepackage{placeins} 
\usepackage{tablefootnote} 
\usepackage[shortlabels]{enumitem}

\usepackage[top=1.5in, bottom=1.5in, left=1in, right=1in]{geometry}
\usepackage[square,numbers,sort&compress]{natbib}
\usepackage[font={small}]{caption}
\usepackage{multirow}
\usepackage{url}

\usepackage{abstract} 

\usepackage{fancyhdr}
\fancyheadoffset{0cm}
\usepackage{titlesec} 
\usepackage{authblk} 
\usepackage{verbatim} 
\usepackage{sectsty}
\allsectionsfont{\boldmath}
\usepackage{orcidlink}
\usepackage{tikz}

\usetikzlibrary{tikzmark}

\def\fwX{\href{http://fwx.pitt.edu}{\texttt{\textsc{fwXmachina}}}}
\def\pT{p_\textrm{\scriptsize T}}
\def\ET{E_\textrm{\scriptsize T}}
\def\ETmiss{\ET^\textrm{miss}}
\def\tkz#1{\tikzmark{#1}}
\def\apf#1{fixed$_{#1}$}
\def\api#1{int$_{#1}$}

\begin{document}

\title{
    \vspace{-36pt}\begin{flushright}{\large PITT-PACC-2409-v1}\end{flushright}\vspace{36pt}
    \textbf{\Large Nanosecond hardware regression trees in FPGA at the LHC}
}

\author[a]{P.\ Serhiayenka}
\author[a,b\,\orcidlink{0000-0002-3878-5873}]{S.~T.~Roche}
\author[a,c\,\orcidlink{0000-0002-7550-7821}]{B.~T.~Carlson}
\author[a\,\orcidlink{0000-0001-7834-328X}]{T.~M.~Hong\thanks{Corresponding author, \href{mailto:tmhong@pitt.edu}{tmhong@pitt.edu}}}

\affil[a]{Department of Physics and Astronomy, University of Pittsburgh}
\affil[b]{School of Medicine, Saint Louis University}
\affil[c]{Department of Physics and Engineering, Westmont College}
\date{\today}

\maketitle
\vfill
\begin{abstract}
\noindent
    We present a generic parallel implementation of the decision tree-based machine learning (ML) method in hardware description language (HDL) on field programmable gate arrays (FPGA). A regression problem in high energy physics at the Large Hadron Collider is considered: the estimation of the magnitude of missing transverse momentum using boosted decision trees (BDT). A forest of twenty decision trees each with a maximum depth of ten using eight input variables of 16-bit precision is executed with a latency of less than 10 ns using $\mathcal{O}$(0.1\%) resources on Xilinx UltraScale+ VU9P---approximately ten times faster and five times smaller compared to similar designs using high level synthesis (HLS)---without the use of digital signal processors (DSP) while eliminating the use of block RAM (BRAM). We also demonstrate a potential application in the estimation of muon momentum for ATLAS RPC at HL-LHC.
\end{abstract}
\vspace{18pt}
    Trigger concepts and systems (hardware and software).
\vfill

\newpage
\tableofcontents

\section{Introduction}

Deep architectures for machine learning (ML) methods continue to empower high energy physics experiments, such as the ATLAS \cite{Aad:2008zzm} and CMS \cite{Chatrchyan:2008aa} experiments at the Large Hadron Collider (LHC) \cite{Evans:2008zzb}. Various simplified approaches for ML designs aimed for trigger systems using field programmable gate arrays (FPGA) exist in the literature. Many use high level synthesis (HLS), where C-like syntax is converted to an RTL circuit using a translator provided by the vendor. The HLS version of neural networks by hls4ml \cite{Duarte:2018ite} gave rise to a body of work in the implementation of artificial intelligence (AI), such as autoencoders and convolutional neural networks \cite{Govorkova:2021utb,Zipper:2023ybp}, on firmware. Similarly, the HLS version of decision trees \cite{Hong:2021snb,Carlson:2022dgb} by \fwX\ (developed by us) also gave rise to an autoencoder \cite{Roche:2024}. There are also non-HLS approaches using hardware description language (HDL) for neural networks \cite{Ghanathe:2017dpm} and decision trees \cite{Summers:2020xiy}, as well as more dedicated applications, e.g., \cite{Ospanov:2021EPJWC,Gonski:2022}.

In this paper, we present a more efficient implementation of decision trees with lower latency. We illustrate the physics potential using the problem in our previous paper~\cite{ATLAS:2020atr}, the estimation of missing transverse momentum ($\ETmiss$) at the LHC.  $\ETmiss$ is an important signature of physics with minimal interaction with the detector materials, including neutrinos and beyond the Standard Model (BSM) particles, including dark matter \cite{Sirunyan:2018owy,ATLAS:2022yvh} and supersymmetry (SUSY) \cite{ATLAS:2019lng,ATLAS:2021kxv,ATLAS:2024umc}. The HL-LHC \cite{Apollinari:2015wtw} upgrades of the trigger systems of ATLAS and CMS experiments \cite{2137107,Zabi:2020gjd} provide an opportunity to implement AI/ML with deeper designs for a variety of applications.

This paper is organized as follows. Section 2 describes the ML training and firmware implementation on FPGA. Section 3 presents the results, with comparisons to previous publications. Appendix A gives the technical details of the subscore adder. Appendix B presents a study, following Ref.~\cite{Ospanov:2022fke}, of the estimation of muon momentum at the High Luminosity LHC (HL-LHC) using hits in the resistive plate chamber (RPC) subdetector in the ATLAS level-0 trigger system.

\section{Method}

The ML training is described followed by the firmware design.
{\textit{T}\textsc{MVA}} \cite{Hocker:2007ht} is used to train a boosted decision tree (BDT) for regression with the ``truth'' $\ETmiss$ as the target variable. The setup follows our previous publication \cite{Carlson:2022dgb} and uses the data samples described therein.

\begin{figure}[!hbtp]
\centering
\includegraphics[width=0.85\textwidth]{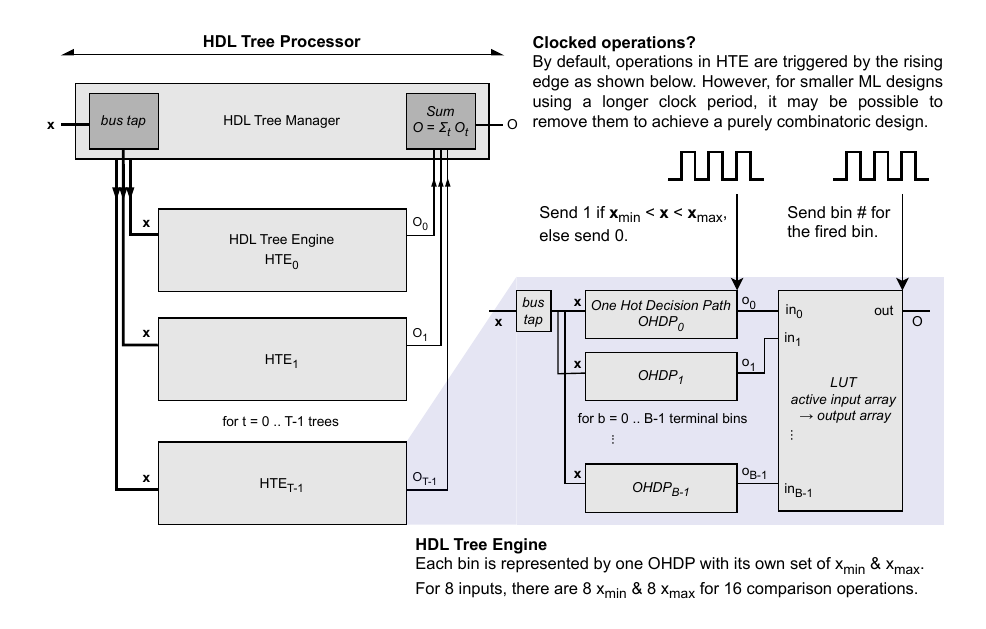}
\caption{
    Block diagram of the VHDL version of the Deep Decision Tree Engine (DDTE). Each tree is represented by HDL Tree Engine (HTE), which are composed of One Hot Decision Paths (OHDP) corresponding to Parallel Decision Paths (PDP) (from Figure 2 of Ref.\ \cite{Carlson:2022dgb}).
    \label{fig:dt}
}
\end{figure}

The firmware design is a VHDL adaptation of the Deep Decision Tree Engine (DDTE) originally written in HLS \cite{Carlson:2022dgb}. In python, \fwX\ program writes VHDL that reflects a given ML configuration from the training. In Figure~\ref{fig:dt}, each tree is represented by a \textsc{HDL Tree Engine} (HTE) and the forest is managed by the \textsc{HDL Tree Manager}. The summing of the subscores is either clocked or combinational; see Appendix A. We note that operations are triggered by the rising edge of the clock, but it may be possible to to remove them for smaller designs and slower clock speeds.

The DDTE proceeds in two steps. The first step bins the inputs and fetches a subscore for each tree represented by HTE in Fig.\ \ref{fig:dt}. For example, four trees produce four subscores. The second step combines the subscores to produce the final score. Division is avoided in the firmware by pre-dividing each subscore by the number of trees. The pre-division is done so that the weighted sum produces the average in the firmware; the firmware only performs the sum and does not divide.

\section{Results}

Results of four configurations are presented, summarized in Table \ref{tab:main}, followed by scaling properties based on one of the configurations. The configurations are
\begin{enumerate}[(1)]
\itemsep=0pt
\item Forty trees with a maximum depth $D=6$ using eight 16-bit input variables at 320 MHz.
\item Ten trees with a $D=8$ and the same as (1) otherwise.
\item Twenty trees with a $D=10$ and the same as (1) otherwise, but using the combinational adder.
The testbench to reproduce this configuration is available online \cite{Serhiayenka:2024}.
\item Hundred trees with a $D=12$ and the same as (1) otherwise.
\end{enumerate}

The first two configurations are identical to those reported previously \cite{Carlson:2022dgb}, which used high level synthesis (HLS) to write the firmware. The latter two demonstrate the speed and efficiency of the new design. Configuration (1) yields a design that is 50\% faster at 25 ns latency and 20 times smaller usage of flip flops, with similar look up table usage using a pipeline adder. Configuration (2) yields a design that is 5 times faster at about 20 ns latency and 7 times smaller usage of flip flops, 5 times smaller usage of look up tables, and eliminates the use of block RAM (BRAM) again using a pipeline adder. Configuration (3) has twice as many trees as Configuration (2), but the design using a combinational adder executes in half the latency with similar resource usage; this result is noted in the abstract. Lastly, Configuration (4) has an order of magnitude more trees as Configuration (2), but the latency grows by less than 10 ns and the resource usage remains modest using a pipeline adder. No digital signal processors (DSP), BRAM, or URAM are used in the four configurations. We note that due to the behavior of the BDT training, the number of bins does not necessarily correspond to a fully populated binary tree, so we define a quantity called ``effective depth'' $d$ so that $2^d = N_\textrm{bin}/N_\textrm{tree}$ represents the number of bins of a fully populated decision tree of depth $d$. For Configurations (2), (3), and (4), the effective depth $d$ is similar at approximately $7$.

\begin{figure}[!hbtp]
\centering
\includegraphics[width=0.48\textwidth]{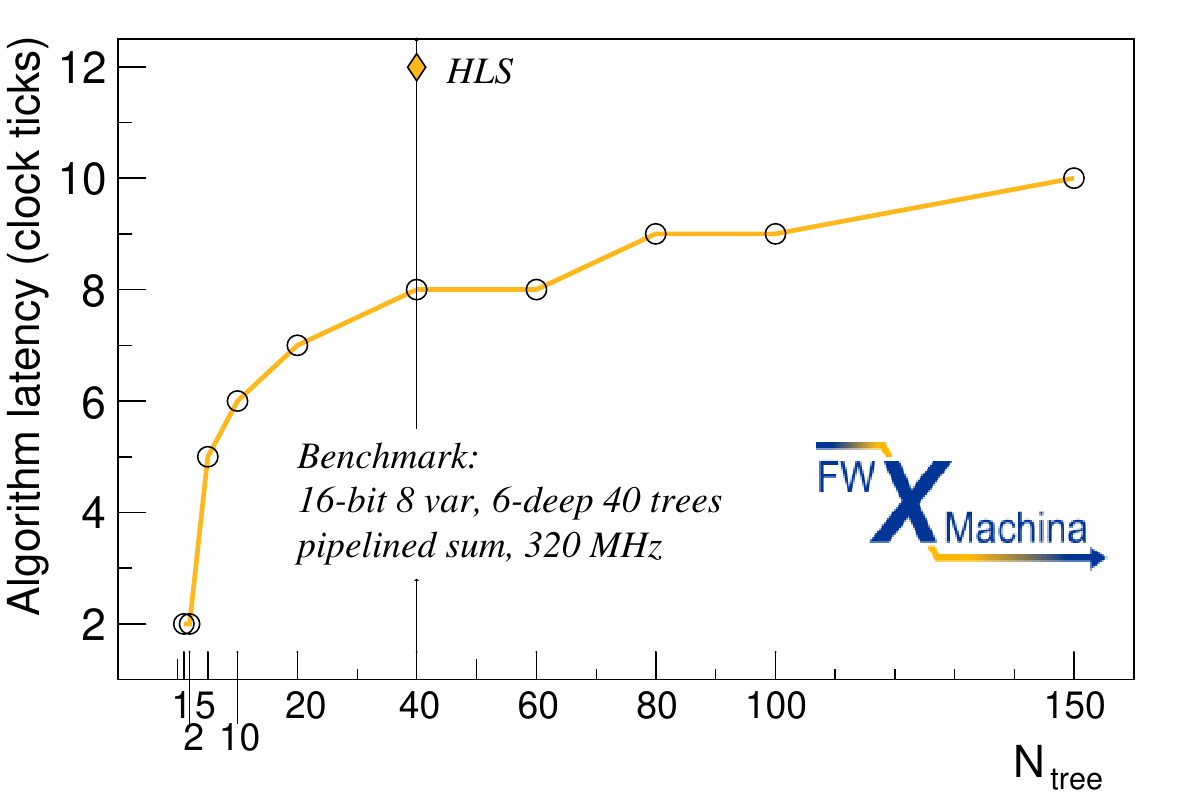}
\caption{
    Algorithm latency scaling vs.\ $N_\textrm{tree}$. The scaling is done with respect to the parameters stated on the plot, corresponding to the benchmark configuration in Table~\ref{tab:main}. Diamond represents the HLS results \cite{Carlson:2022dgb}.
    \label{fig:scaling1}
}
\end{figure}

Scaling properties are studied by changing one parameter at a time with respect to the first configuration described above as the benchmark. The algorithm latency for Configuration 1 is eight clock ticks. The latency can be as low as two clock ticks, but it grows logarithmically to ten clock ticks for 150 trees as shown in Fig.\ \ref{fig:scaling1}. The logarithmic dependence is due to the adder being arranged in a binary tree structure: two subscores are iteratively summed until a final score is reached. In contrast to the pipelined adder, the combinational adder always has a latency of 2 clock ticks. However, it has limited use case as the model has to be relatively simple. For further discussion about the adder, we refer to Appendix \ref{sec:clock}. Lastly, the interval period between successive inputs is one clock tick for all tested configurations, so the firmware can take in an input at every clock tick and generate an output after the algorithm latency.

\begin{figure}[!hbtp]
\centering
\includegraphics[width=0.48\textwidth]{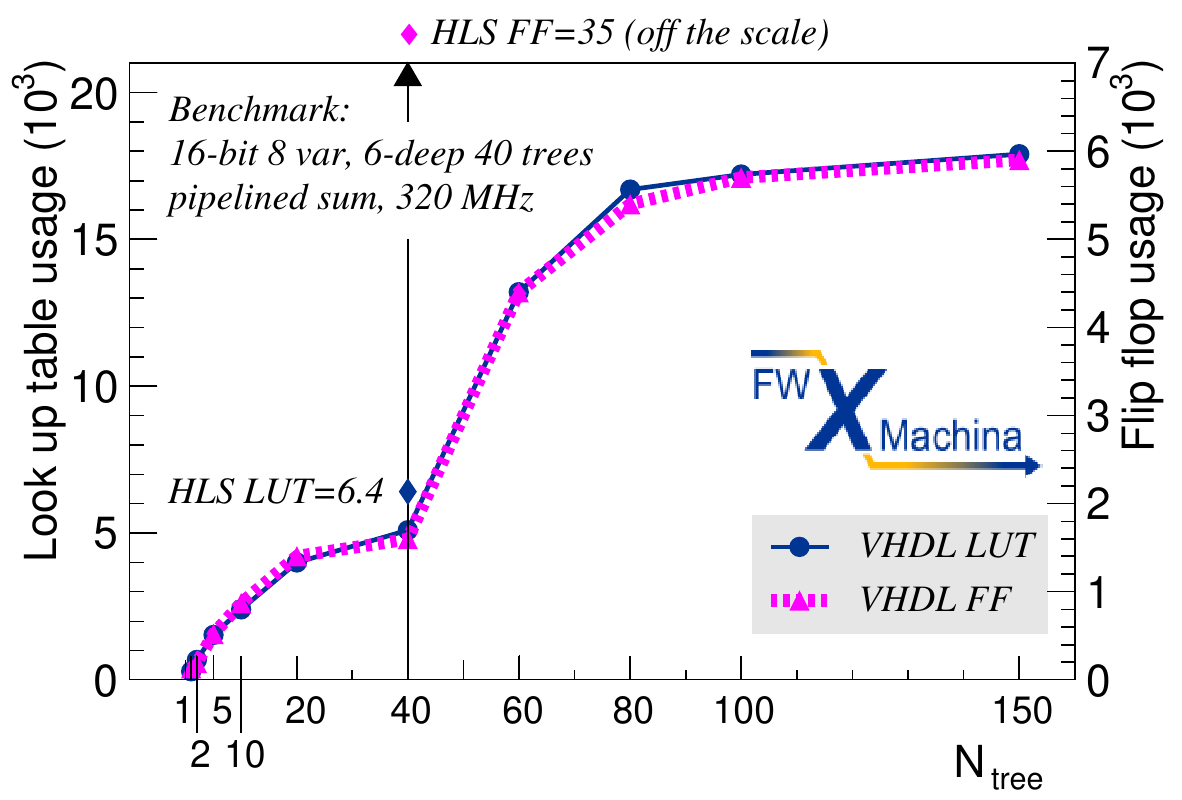}\hspace{0.02\textwidth}%
\includegraphics[width=0.48\textwidth]{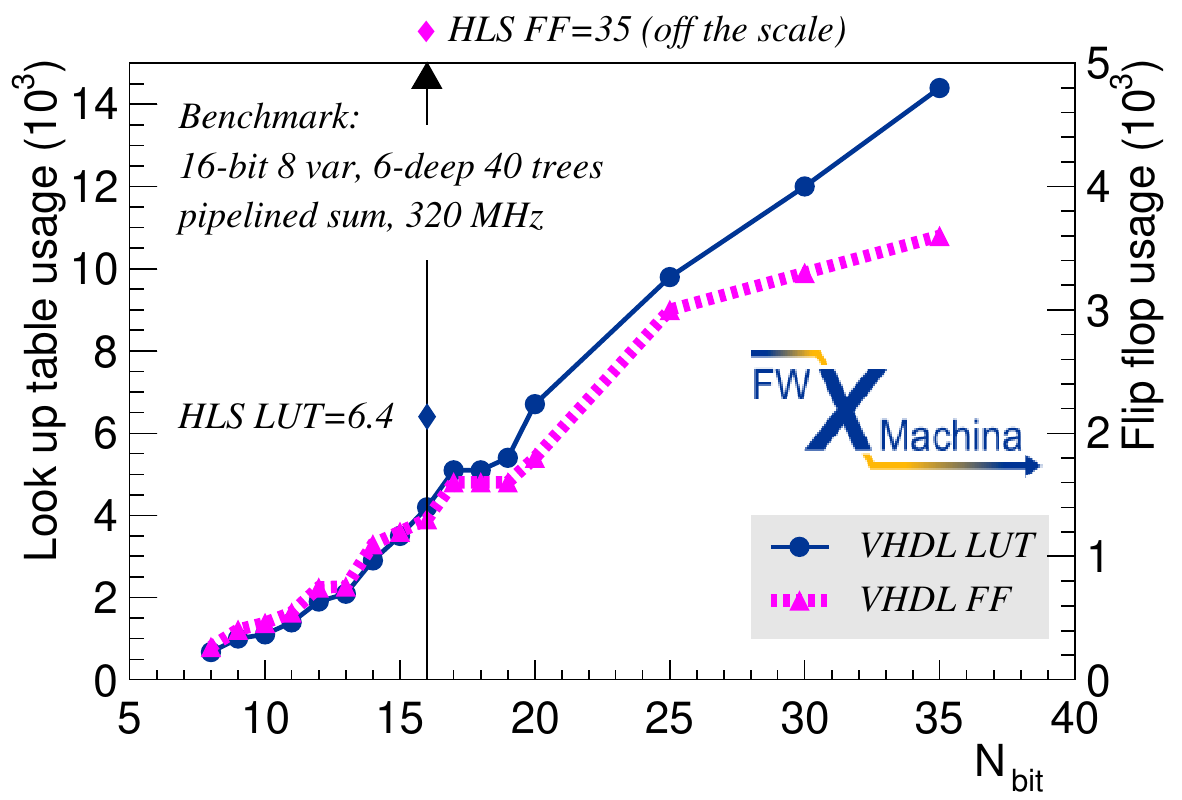}%
\caption{
    Resource usage scaling vs.\ $N_\textrm{tree}$ (left) and $N_\textrm{bit}$ (right). On each plot, look up table usage is shown in circles with the scale on the left side and flip flop usage is shown in triangles with the scale on the right side. The scaling is done with respect to the parameters stated on the plot, corresponding to the benchmark configuration in Table~\ref{tab:main}. Diamond represents the HLS results \cite{Carlson:2022dgb}, with flip flop usage being off the scale and noted as such with the arrow.
    \label{fig:scaling2}
}
\end{figure}

Resource usage is considered as a function of the number of trees ($N_\textrm{tree}$) in the forest and the number of bits ($N_\textrm{bit}$) of each input variable. The dependence of look up tables and flip flops on the  the number of trees is expected to be linear as the combinational logic scales linearly. The left plot in Fig.\ \ref{fig:scaling2} shows this to be roughly true until about eighty trees, but the dependence seems to flatten out afterwards. For the number of bits, we observe a linear relationship for both LUT and FF on the right plot of Fig.\ \ref{fig:scaling2}, as expected. The separate study for the combinational adder was not done, but we anticipate that usage results would be similar to the pipeline design because only the clock is removed.

\begin{table}[htbp!]
\caption{
    FPGA results and comparison with Refs.\ \cite{Summers:2020xiy,Hong:2021snb,Carlson:2022dgb}. All results in the table uses the same FPGA model Xilinx Ultrascale+ VU9P (vu9p-flgb2104-2L-e) with the following available resources  $1.2$\,M LUT, $2.4$\,M FF, $6.8$\,k DSP, and $4.3$\,k BRAM. Effective depth $d$ is defined as so that $2^d = N_\textrm{bin}/N_\textrm{tree}$.
    \label{tab:main}
    }
\centering
{\small
\begin{tabular}{
    p{1.120in}
    p{0.275in}
    p{0.275in}
    p{0.465in}
    p{0.465in}
    p{0.465in}
    p{0.465in}
    p{0.465in}
    p{0.465in}
}
\hline
Goal
& \multicolumn{1}{l}{5 classif'n\!\!}
& \multicolumn{1}{l}{2 classif'n\!\!\!\!}
& \multicolumn{2}{l}{$\ETmiss$ regression}
& \multicolumn{4}{l}{$\ETmiss$ regression\dotfill}
\\
Reference
& \cite{Summers:2020xiy}
& \multicolumn{1}{l}{\cite{Hong:2021snb}}
& \multicolumn{2}{l}{\cite{Carlson:2022dgb}\dotfill}
& \multicolumn{4}{l}{This paper\dotfill}
\\
\hline
Setup \\
\quad Design           &VHDL    &HLS    &HLS       &HLS       &VHDL      &VHDL      &VlDL    &VHDL    \\
\quad Sum strategy     &-       &-      &-         &-         &pipeline  &combin.   &combin. &pipeline\\
\quad Parallelize      &-       &cutwise&pathwise  &pathwise  &pathwise  &pathwise  &pathwise&pathwise\\
\quad Clock (MHz)      &250     &320    &320       &320       &320       &320       &200     &320     \\
\quad Bit precision    &\apf{18}&\api{8}&\api{16}  &\api{16}  &\api{16}  &\api{16}  &\api{16}&\api{16}\\
\quad $N_\textrm{var}$ &16      &4      &8         &8         &8         &8         &8       &8       \\
\quad $N_\textrm{tree}$&100     &100    &40        &10        &40        &10        &20      &100     \\
\quad Max.\ depth $D$  &4       &4      &6         &8         &6         &8         &10      &12      \\
\quad $N_\textrm{bin}$ &-       &-      &1.7\,k    &1.4\,k    &1.7\,k    &1.4\,k    &2.9\,k  &15.7\,k \\
\quad Effective\,depth\,$d$\!\!
                       &-       &-      &5.4\tkz{a}&7.2\tkz{b}&5.4\tkz{c}&7.2\tkz{d}&7.2     &7.3 
\vspace{6pt}
\\
\\
\quad Notable&&
& \multicolumn{2}{l}{\qquad\quad identical}
& \multicolumn{2}{l}{\quad identical}
\begin{tikzpicture}[remember picture, overlay]
  \draw [<-,solid] ([yshift=-1ex]pic cs:c) [bend left] to ([yshift=-1ex]pic cs:a);
  \draw [<-,dashed] ([yshift=-1ex]pic cs:d) [bend left] to ([yshift=-1ex]pic cs:b);
\end{tikzpicture}
& slower clock
& larger forest
\\
\hline
Results \\
\quad LUT              &96\,k   &1\,k      &6.4\,k   &75\,k    &5.1\,k   &10\,k    &15.5\,k &38\,k   \\
\quad FF               &43\,k   &0.1\,k    &35\,k    &24\,k    &1.6\,k   &4.7\,k   &6.6\,k  &19.4\,k \\
\quad DSP              &0       &2         &0        &0        &0        &0        &0       &0       \\
\quad BRAM             &0       &5.5       &0        &10       &0        &0        &0       &0       \\
\quad URAM             &-       &0         &0        &0        &0        &0        &0       &0       \\
\quad Latency (ns)     &52\,ns  &9.375\,ns &38\,ns   &119\,ns  &25\,ns   &19\,ns   &10\,ns  &28\,ns  \\
\qquad $''$\qquad(tick)&13      &3         &12       &38       &8        &6        &2       &9       \\
\quad Interval~\,(tick)&1       &1         &1\tkz{e} &1\tkz{f} &1\tkz{g} &1\tkz{h} &1       &1
\vspace{6pt}
\\
\\
\quad Notable &&
& \multicolumn{2}{l}{\qquad benchmark}
& \multicolumn{2}{l}{}
& in\,abstract 
& 
\\
\hline
\begin{tikzpicture}[remember picture, overlay]
  \draw [<-,solid] ([yshift=-1ex]pic cs:g) [bend left] to ([yshift=-1ex]pic cs:e);
  \draw [<-,dashed] ([yshift=-1ex]pic cs:h) [bend left] to ([yshift=-1ex]pic cs:f);
\end{tikzpicture}
\end{tabular}
}
\end{table}

The improved ML architecture presented in this paper may be of use in constrained environments such as in the trigger systems of ATLAS \cite{2137107} and CMS \cite{Zabi:2020gjd}, which are being upgraded for the High Luminosity LHC (HL-LHC) to account for the substantial increase in the average number of simultaneous proton collisions per bunch crossing, reaching values of up to $200$ \cite{Apollinari:2015wtw}.

\FloatBarrier 

\section*{Acknowledgments}

We thank Connor Marsh for assistance with FPGA data collection. We thank Elliot Lipeles for discussions regarding tails of neural network distributions. We thank Yuvaraj Elangovan and Isabelle Taylor of the Electronics Shop. Work performed in the University of Pittsburgh Dietrich School Electronics Shop Core Facility (RRID:SCR\_025113) and services and instruments used in this project were graciously supported, in part, by the University of Pittsburgh. TMH was supported by the US Department of Energy [award no.\ DE-SC0007914]. BTC was supported by the National Science Foundation [award no.\ 2209370]. STR was supported by the Emil Sanielevici Scholarship.

\appendix

\section{Adder details}
\label{sec:clock}

\begin{figure}[!hbtp]
\begin{center}
\includegraphics[width=0.50\textwidth]{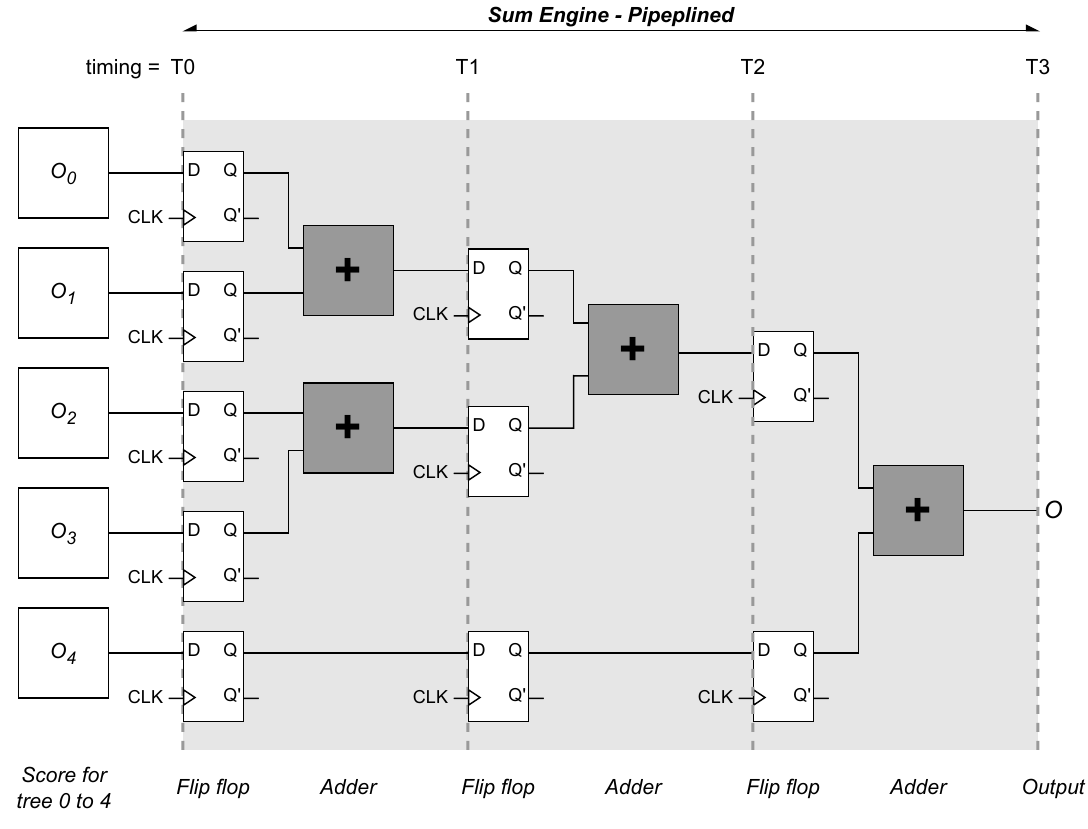}%
\includegraphics[width=0.50\textwidth]{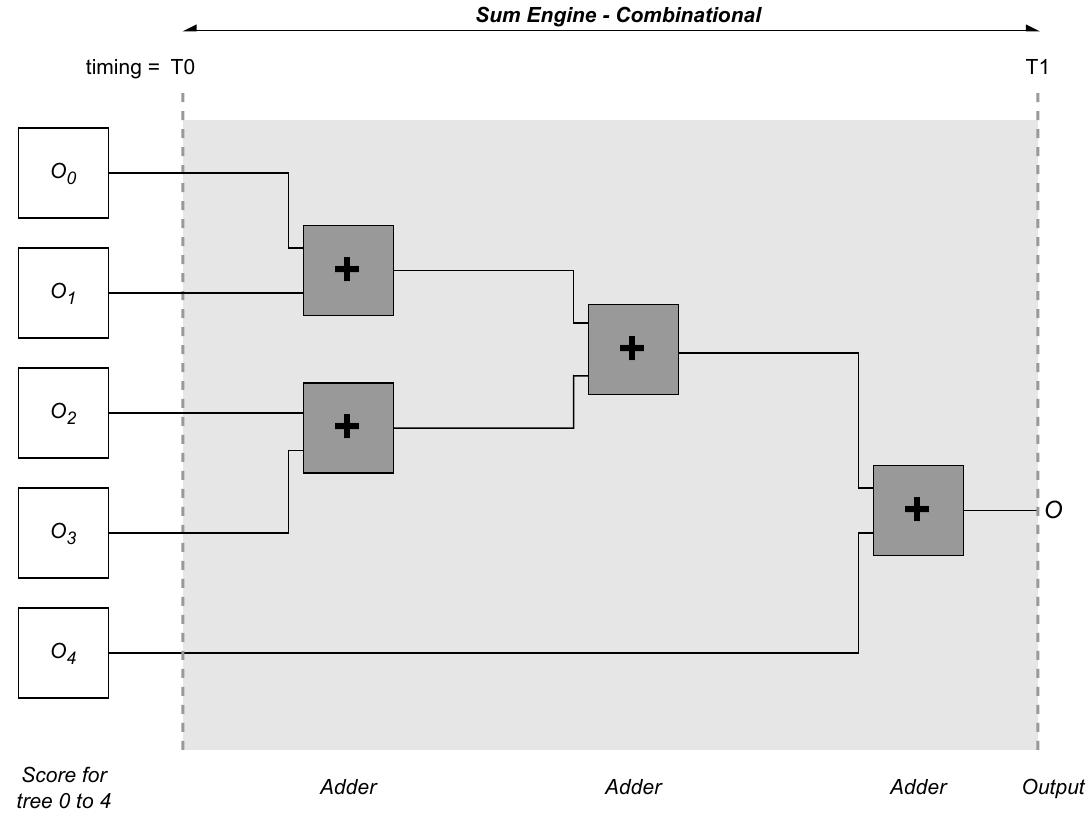}
\end{center}
\caption{
    Two adder designs using pipeline (left) and combinational logic (right) for the sum block in Fig.~\ref{fig:dt}.
    \label{fig:hte}
}
\end{figure}

Two firmware designs are implemented to execute the addition of the tree subscores. The first is combinational, which organizes addition into a tree structure, independent of the clock, as shown in the left diagram of Fig.~\ref{fig:hte}. The second technique is a pipeline, which also organizes addition into a tree structure but with clocks inserted in between each level of the adder. Flip flops are used to separate the addition operations into distinct, clocked stages as shown in the right of Fig.~\ref{fig:hte}. This results in an adder that is dependent on the clock. Only one level of the adding process is computed during each clock tick, which results in as many clock ticks are there are levels.

The advantage of the combinational adder is speed. Once the adder receives a set of subscores, it will begin to compute the sum immediately from start to finish. The disadvantage is that it cannot support a large number of trees without timing violations, as it assumes that all the trees' scores may be added within a single clock tick. Because the pipeline adder distributes the subscore addition process into steps, assuming all other parameters are equal, it can operate for a larger number of trees than the combinational. Though, the pipeline adder does have upper limits of operation, these limits are much higher than those of the combinational. The disadvantage of the pipeline adder is that it will have a higher latency compared to the same size design using a combinational adder. While the combinational latency is always 2 ticks, the pipeline latency is represented by 2 plus the ceiling integer of $\log_2$ of the number of trees, i.e., latency in clock ticks $= 2 + \lceil \log_2 N_\textrm{trees}\rceil$. Therefore, combinational may be desired with pipeline reserved for scenarios where timing violations occur.

\section{Muon momentum for ATLAS RPC at HL-LHC}\label{sec:muon}

We present an application of ML regression for muon momentum estimation in the ATLAS level-0 trigger system for HL-LHC upgrade, which detects muons with resistive plate chamber (RPC) detector in the barrel section \cite{Cardarelli:1988rpc,ATLAS:2021pft}. There are three cylindrical $2\,\textrm{cm}$-wide RPC layers; the trajectory of a muon passing between them can be used to determine its charge and $\pT$ as they are immersed in a toroidal magnetic field.\footnote{An interesting and related problem in ML-based FPGA muon tracking was done for thin gap chambers (TGC) detectors located outside of the toroidal magnetic field \cite{Sun:2022bxx}, but is beyond the scope of this paper.}  Reference~\cite{Ospanov:2021EPJWC} presents an HDL implementation of neural network (NN) regression on FPGA to estimate the $\pT$ using hit information. We apply our BDT to this problem following the detailed description by the same authors \cite{Ospanov:2022fke}.

\begin{figure}[!hbtp]
\centering
\includegraphics[width=0.495\textwidth]{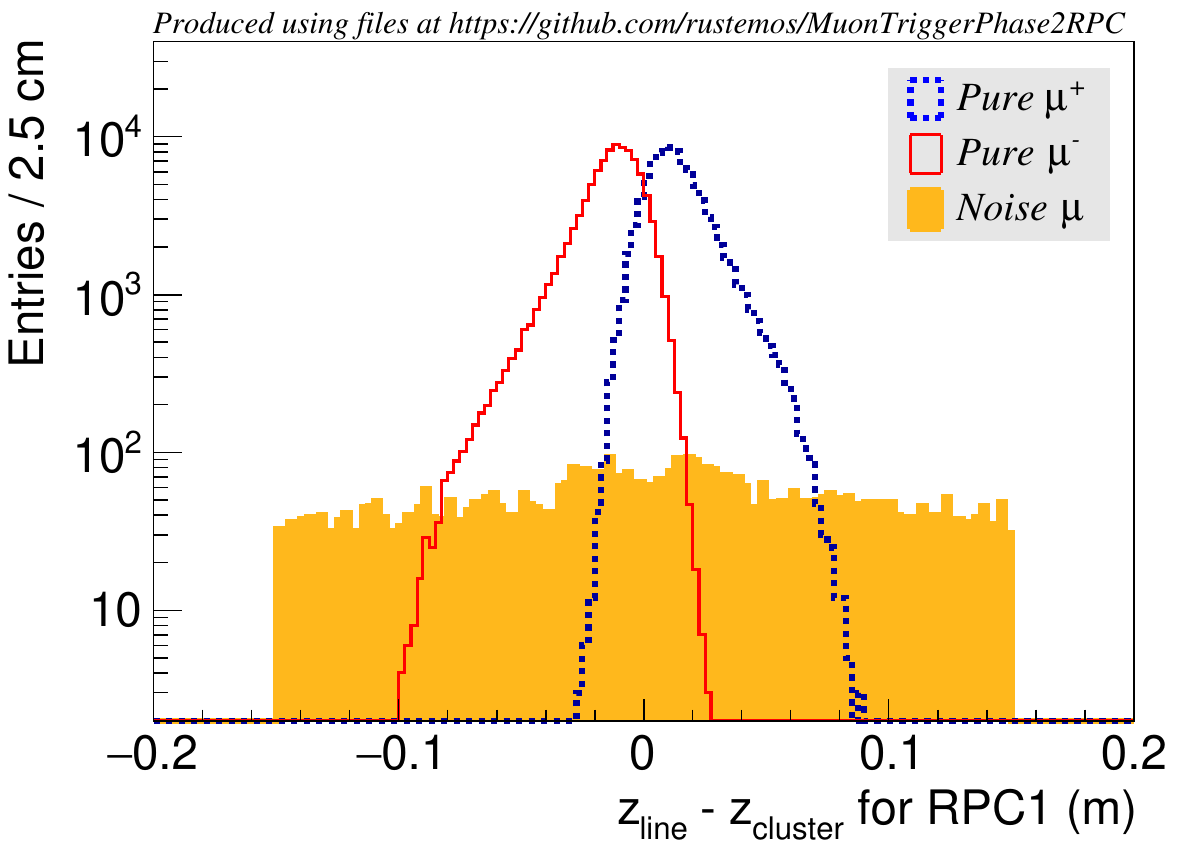}\hspace{0.005\textwidth}%
\includegraphics[width=0.495\textwidth]{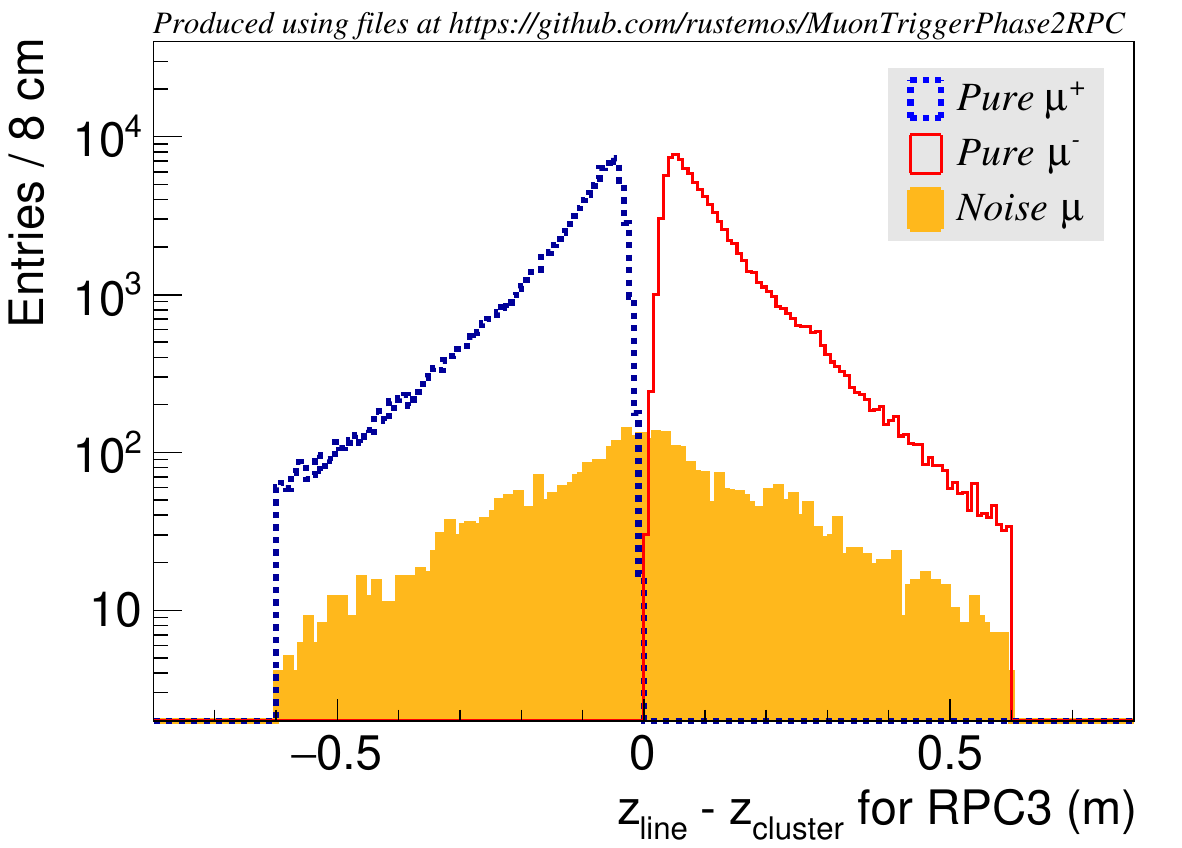}
\caption{
    Distribution of the differences of $z$ coordinate between the impact point of the seed line and the cluster position in the RPC1 (left) and RPC3 (right). Muons are separated by charge and noise fakes. The sample is produced using the code available at Ref.~\cite{MuonTriggerPhase2RPC}. 
    \label{fig:muons_input}
}
\end{figure}

A sample of 200\,k muon candidates were generated in a idealized RPC system of uniformly distributed $\pT$ between $3$ and $30\,\textrm{GeV}$ using the publicly available code in Ref.~\cite{MuonTriggerPhase2RPC}. Half of the sample is positively charged and the other half negatively charged. A small fraction of the sample ($0.1\%$) consists of noise hits from non-muon detector background. For each muon candidate, the $z$-coordinate of the clusters at RPC1, RPC2, and RPC3 are recorded, labeled as $z_1$, $z_2$, and $z_3$, respectively. The three input variables to the regression algorithm are $z_\text{2}$, $\Delta z_3$, and $\Delta z_1$ with the latter two defined as the coordinate relative to $z_2$ as $z_3-z_2$ and $z_2-z_1$, respectively. The two $\Delta z$ distributions are shown in Fig.~\ref{fig:muons_input} and they qualitatively agree with Figure 3 of Ref.~\cite{Ospanov:2022fke}.

The ML setup is described for the BDT as well as for the NN to reproduce results of Ref.~\cite{Ospanov:2021EPJWC}. For both methods, the sample is split evenly into training and testing sets.
For the BDT, a forest of $50$ trees at a maximum depth of $7$ is trained using adaptive boosting (AdaBoost) trained the untransformed variables and set to target the product of charge and $\pT$: $q{\cdot}\pT$. We note that linear transformations on input variables may incur additional latency and resource usage so we avoid them. As done in our previous papers, floating point values are used for training and $16$-bit integers are used for evaluation.

For the NN, we follow the process as was done in Ref.~\cite{MuonTriggerPhase2RPC}. The network is trained on linearly transformed variables---i.e., scaled and shifted to have a null mean and unit standard deviation---and set to target the charge divided by its $\pT$: $q/\pT$. The NN is trained with the default setup of Ref.~\cite{Ospanov:2021EPJWC}; floating point values are used in training. Floating point values are used for evaluation rather than the 16-bit fixed point numbers used in \cite{Ospanov:2021EPJWC} to simplify the analysis without putting its results at a possible disadvantage.

\begin{figure}[!hbtp]
    \centering
    \includegraphics[width=0.5\textwidth]{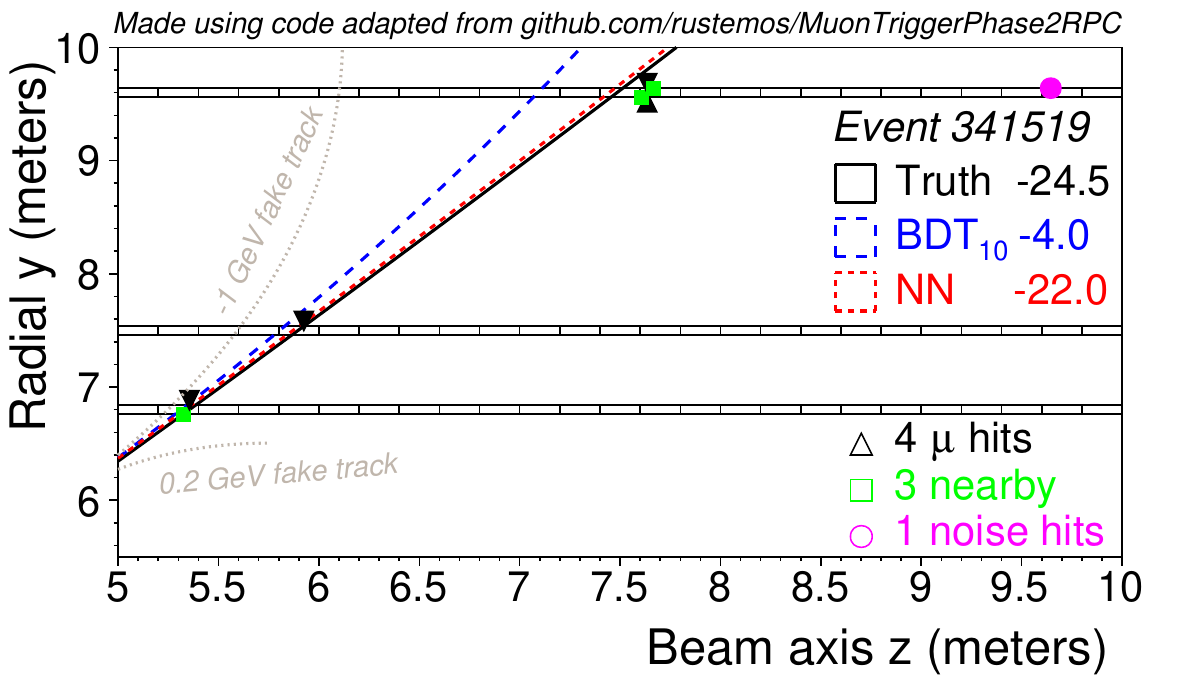}%
    \includegraphics[width=0.5\textwidth]{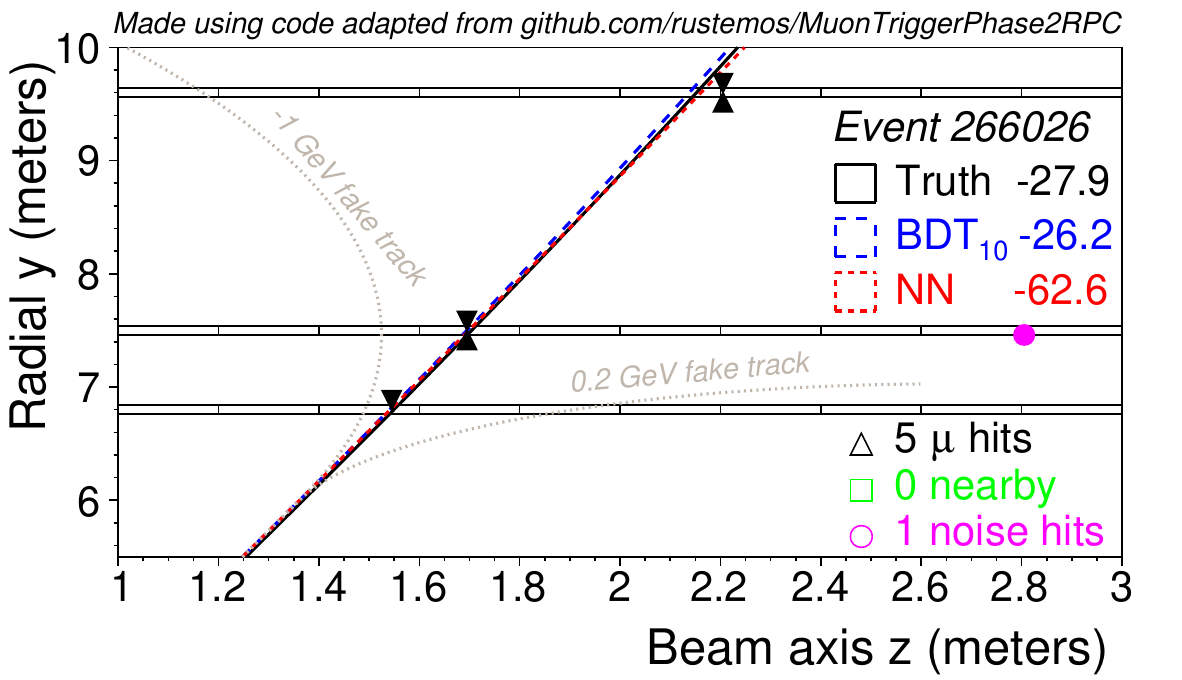}
    \caption{
    Event displays showing the $y$-$z$ projection of the RPC hits with projected tracks using the $\pT$ starting at the origin. The top plot shows the case where NN gives a good estimate at $r=0.9$ and BDT an erroneous one at $r=0.2$, where $r\equiv(q\cdot\pT)^\textrm{reco}/(q\cdot\pT)^\textrm{truth}$. The bottom plot shows the case where BDT gives a good estimate at $r=0.9$ and NN an erroneous one at $r=2.2$. As a sanity check, we overlay two fake tracks (unrelated to the hits in the display) with $q\cdot\pT=-1$ GeV and $0.2$ GeV to visually demonstrate the curvature dependence on $q\cdot\pT$. Muon hits and nearby hits, excluding noise hits, are used to compute the BDT and NN scores. The BDT configuration for these event displays uses 100 trees with a maximum depth of 10.
    \label{fig:muon_display}
    }
\end{figure}

The physics performance is evaluated by considering the resolution of the estimated charge times $\pT$, defined as the ratio of the reconstructed and the the truth value, i.e., $r\equiv(q\cdot\pT)^\textrm{reco}/(q\cdot\pT)^\textrm{truth}$. The ratio is chosen as metric for comparing the BDT to the NN because it can be evaluated using a sample of simulated muons. Two events displays are shown in Fig.~\ref{fig:muon_display} to show a case where the NN gives a good estimate whereas the BDT gives an erroneous one, and vice versa.
The $\pT$ value of the tracks are relatively high enough where the curvature is difficult to see by eye, so two fake tracks with opposite charge and very low $\pT$ are shown to demonstrate the dependence on the curvature from $q\cdot\pT$. Lastly, we note that efficiency turn-on curves are not shown, as a realistic sample of the background composition would be needed to determine the trigger threshold value corresponding to a given background rate.

\begin{figure}[!hbtp]
    \centering
    \includegraphics[width=0.80\textwidth]{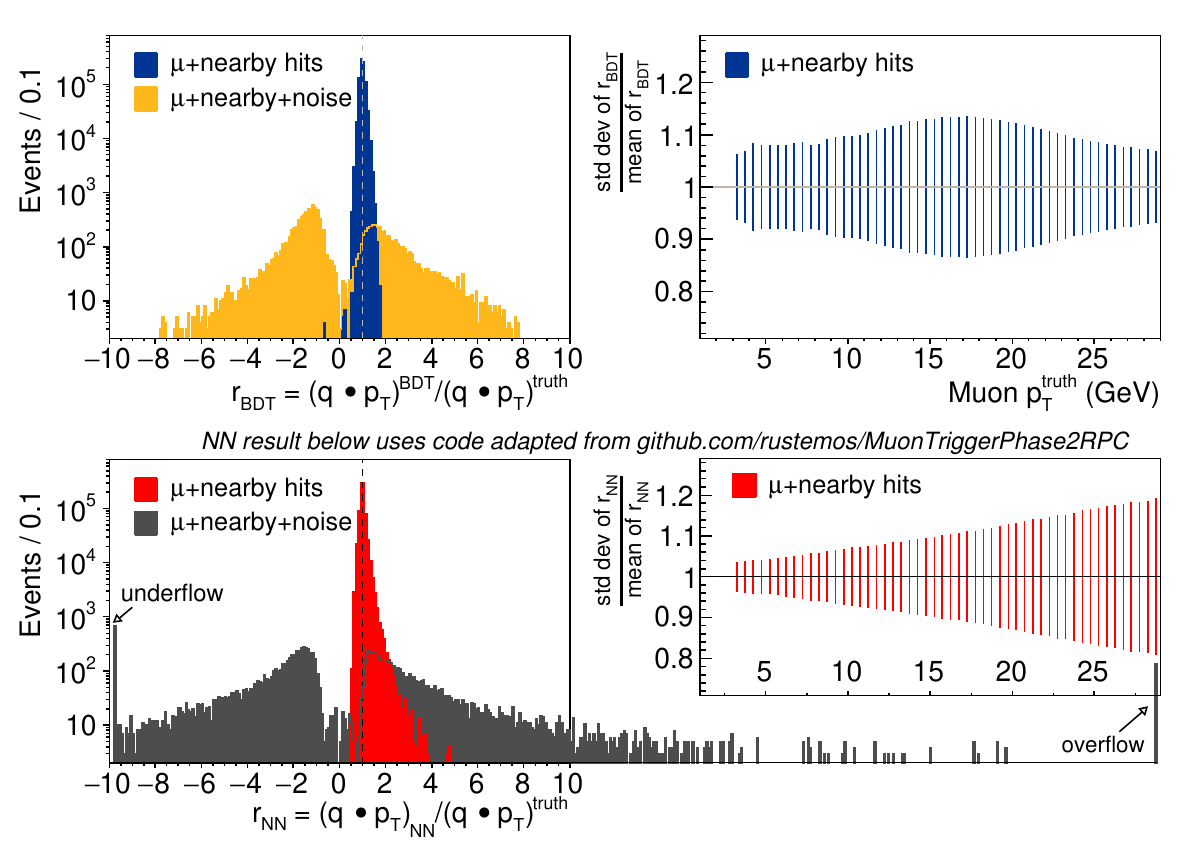}
    \caption{
    (Left column) Distributions of the resolution of muon $\pT$ defined as the ratio with respect to truth $\pT$. (Right column) Profile plots of the standard deviation normalized to the mean defined as $s_r\equiv\textrm{stddev}_r/\textrm{mean}_r$ is shown in slices of truth $\pT$. (Top row) Results are using the BDT configuration corresponding to Table~\ref{tab:muon}. (Bottom row) Results are using the neural network configuration described in the text. For these plots, nearby hits are included with the ``hits from $\mu$.''
    \label{fig:muon_res}
    }
\end{figure}

The physics results are summarized by the $r$ distributions in the left column of Fig.~\ref{fig:muon_res}. The resolution $r$ is shown twice in each plot, once using hits only from the muon and nearby hits and once using those hits and the noise seed. For the results without noise hits, $r$ is peaked at unity for both BDT and NN with standard deviation of $0.116$ and $0.130$, respectively. The profile plot of the standard deviation normalized by the mean defined as $s_r\equiv\textrm{stddev}_r/\textrm{mean}_r$ shows different patterns for BDT and NN. The BDT shows a relatively flat dependence of about $s=10\%$ for the full $\pT$ range. In contrast, the NN shows a smaller value of about $s=4\%$ at 3 GeV that grows linearly with $\pT$ to about $s=20\%$ at 30 GeV. For the results with noise hits, $r$ is peaked at both $1$ and $-1$, which signifies that the charge designation is incorrect at least half of the time, with a much broader spread of its distribution. The NN distribution with noise hits seem to be broader than the one for BDT, so the graphical axis is expanded with underflow and overflow. We note that these studies rely on the mock-up of the ATLAS RPC, so a more realistic considerations could improve the results.

\begin{figure}[!hbtp]
    \centering
    \includegraphics[width=0.98\textwidth]{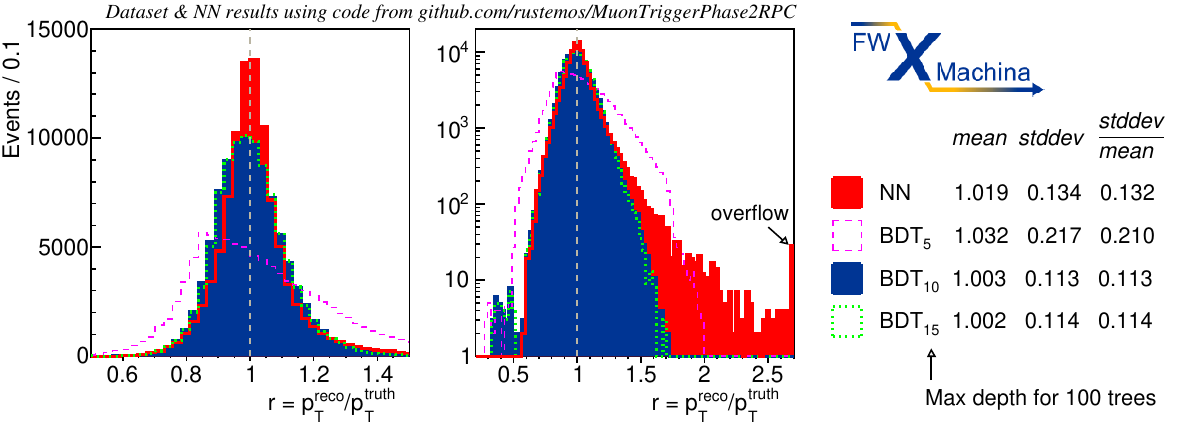}
    \caption{
    Distribution of the resolution of muon $\pT$ for NN and BDT of varying maximum depths with 100 trees. The same distribution is shown twice with linear (left) and logarithmic axes (right).
    \label{fig:muon_res_logy}
    }
\end{figure}

The trade-off between physics and engineering performance is a consideration when implementing the algorithm on FPGA. A simpler ML design would be faster and smaller at the cost of worse resolution $r$. We studied the dependence by varying the maximum depth $D$---5, 10, 15---for hundred trees. As done above, we consider $s_r$ for each configuration. The NN value is $s_r=0.13$ and is smaller than $s_r=0.21$ obtained for BDT at a relatively shallow $D=5$. A deeper configuration of $D=10$ yields a competitive value of $s_r=0.11$, but no further improvement is seen for even deeper configuration as $D=15$ yields the same value. The full table of values are given in Fig.~\ref{fig:muon_res_logy}. The figure shows the distributions in linear and logarithm scales. The former highlights the relative sharpness of the NN near unity with respect to BDT with $D=10$, whereas the latter shows the longer tail of the NN. One typical event where $r=0.2$ for BDT$_{10}$ and another events where $r=2.2$ for NN are shown in the event displays of Fig.~\ref{fig:muon_display}.

Finally, the FPGA results are presented for BDT of thirty trees with a maximum depth of seven. This combination is chosen to be close to the physics performance of the neural network. The resource utilization is about 11\,k of look up tables and 4\,k of flip flops are used with no DSP, BRAM, or URAM usage. The algorithm latency is 22\,ns, which corresponds to 7 clock ticks at 320 MHz. The interval between successive inputs is 1 clock tick. The look up table and flip flop usage is comparable to the NN, but our design does not use DSP due to lack of multiplication operations. We note that the NN timing results are from simulated values whereas our results are from the actual values measured on the board, so further differences may arise with NN measurements after implementation.

\begin{table}[hbtp!]
\caption{
    FPGA results for muon momentum estimation for ATLAS RPC at HL-LHC.
    \label{tab:muon}
}
\centering
{\small
\begin{tabular}{llllll}
\hline
Method
& \multicolumn{2}{l}{Neural network \cite{Ospanov:2021EPJWC}}
& \multicolumn{2}{l}{BDT [this work]}
\\ \hline
Setup
\\ \quad Design            &\multicolumn{2}{l}{Verilog}                  &\multicolumn{2}{l}{VHDL}
\\ \quad Xilinx model      &\multicolumn{2}{l}{Kintex UltraScale XCKU060}&\multicolumn{2}{l}{Virtex UltraScale+ VU9P}
\\ \quad Clock             &\multicolumn{2}{l}{320 MHz}                  &\multicolumn{2}{l}{320 MHz}
\\ \quad Variable precision&\multicolumn{2}{l}{\apf{16}}                 &\multicolumn{2}{l}{\api{16}}
\\ \quad Input $N_\textrm{var}$ &\multicolumn{2}{l}{3}                   &\multicolumn{2}{l}{3}
\\ \quad Target variable   &\multicolumn{2}{l}{$q/p_T$}                  &\multicolumn{2}{l}{$q\cdot p_T$}
\\ \quad ML method         &\multicolumn{2}{l}{Neural network}           &\multicolumn{2}{l}{Boosted decision tree}
\\ \quad NN configuration  &\multicolumn{2}{l}{3 hidden layers, fully connected} &\multicolumn{2}{l}{-}
\\ \quad                   &\multicolumn{2}{l}{30 nodes per layer}       &\multicolumn{2}{l}{-}
\\ \quad                   &\multicolumn{2}{l}{ReLU activation function} &\multicolumn{2}{l}{-}
\\ \quad BDT configuration &\multicolumn{2}{l}{-}                        &\multicolumn{2}{l}{30 decision trees}
\\                         &\multicolumn{2}{l}{-}                        &\multicolumn{2}{l}{Maximum depth of 7}
\\                         &\multicolumn{2}{l}{-}                        &\multicolumn{2}{l}{Boosting strategy: adaptive}
\\ \quad                   &\multicolumn{2}{l}{-}                        &\multicolumn{2}{l}{Sum strategy: pipeline}
\\ \hline               
Results                 
\\ \quad LUT               &4659 &(1.4\%)                                &\multicolumn{2}{l}{11.3\,k}
\\ \quad FF                &7370 &(1.1\%)                                &\multicolumn{2}{l}{3.8\,k}
\\ \quad DSP               &157  &(5.7\%)                                &\multicolumn{2}{l}{0}
\\ \quad BRAM              &\multicolumn{2}{l}{Not reported}             &\multicolumn{2}{l}{0}
\\ \quad URAM              &\multicolumn{2}{l}{Not reported}             &\multicolumn{2}{l}{0}
\\ \quad Latency           &122\,ns &(simulated)                         &22\,ns   &(measured)
\\ \qquad $''$             &39\,ns  &(simulated)                         &7\,ticks &(measured)
\\ \quad Interval          &8\,tick &(simulated)                         &1\,tick  &(measured)
\\ \hline
\end{tabular}
}
\end{table}

\FloatBarrier 

\clearpage

\addcontentsline{toc}{section}{References}
\bibliographystyle{JHEP}

\end{document}